\documentclass[12pt]{article}
\usepackage{graphicx}
\begin{document}

\title{Effect of nuclear periphery on nucleon transfer in peripheral collisions}

\author{Martin Veselsky\\
\\
Institute of Physics, Slovak Academy of Sciences,\\
Dubravska cesta 9, Bratislava, Slovakia\\
% Tel: +421-2-59410532, Fax: +421-2-5477-6085\\
e-mail: fyzimarv@savba.sk\\
\\
and\\
\\
G.A. Souliotis\\
\\
Cyclotron Institute, Texas A\&M University, \\ 
College Station, USA
}

\date{}

\maketitle

\begin{abstract}
A comparison of experimental heavy residue cross sections 
from the reactions $^{86}$Kr+$^{64}$Ni,$^{112,124}$Sn with 
the model of deep-inelastic transfer ( DIT ) is carried 
out. A modified expression for nucleon transfer probabilities 
is used at non-overlapping projectile-target configurations, 
introducing  a dependence on isospin asymmetry at the nuclear periphery. 
The experimental yields of neutron-rich nuclei close to the projectile are 
reproduced better and the trend deviating from the bulk isospin equilibration 
is explained. For the neutron-rich products further from the projectile, 
originating from hot quasiprojectiles, the statistical 
multifragmentation model reproduces the mass distributions 
better than the model of sequential binary decay. 
In the reaction with proton-rich target $^{112}$Sn 
the nucleon exchange appears to depend on isospin asymmetry 
of nuclear periphery only when the surface separation 
is larger than \hbox{0.8 fm} due to the stronger 
Coulomb interaction at more compact di-nuclear configuration. 

\end{abstract}

\section*{Introduction}

Nucleus-nucleus collisions in the Fermi energy domain exhibit a large variety 
of contributing reaction mechanisms and reaction products ( see e.g. 
\cite{MVNPA}, \cite{GSKrNi}, \cite{GSIso}, \cite{MVSnAl} ) and offer  
the principal possibility 
to produce mid-heavy to heavy neutron-rich nuclei in very peripheral collisions. 
In the reactions of massive heavy ions such as $^{86}$Kr+$^{124}$Sn 
\cite{GSKrSn} and $^{124}$Sn+$^{124}$Sn \cite{GSSnSn}, an enhancement 
was observed over the yields expected in cold fragmentation 
which is at present the method of choice to produce 
neutron-rich nuclei. Further enhancement of yields of n-rich nuclei 
was observed in the reaction $^{86}$Kr+$^{64}$Ni \cite{GSKrNi} in 
the very peripheral collisions, thus pointing to the possible 
importance of neutron and proton density profiles 
at the projectile and target surfaces. 

In this article, we present an investigation of the reaction mechanism 
of the very peripheral nucleus-nucleus collisions with an emphasis on 
the possible role of different proton and neutron densities at the 
surface as a factor influencing the process of nucleon exchange. 
When nuclei with different neutron to proton ratio collide peripherally 
and nucleon exchange takes place, one observes equilibration 
of the N/Z-ratio. An isoscaling study was carried out 
on the reactions $^{86}$Kr+$^{112}$Sn and $^{86}$Kr+$^{124}$Sn at 25 AMeV 
\cite{GSIso} and a correlation of the isoscaling parameter with 
isospin equilibration was observed \cite{GSNZPLB}. 
Peripheral nucleus-nucleus collisions can be described theoretically 
using the model of deep-inelastic 
transfer \cite{DIT} in combination with an appropriate model of de-excitation. 
Very good description of experimental observables was obtained 
\cite{GSKrNi,GSKrSn,GSSnSn,MVSiSn} 
using the DIT model of Tassan-Got \cite{TG} 
and the de-excitation codes SMM \cite{SMM} and GEMINI \cite{GEM}.  
SMM represents the statistical model of multifragmentation ( SMM ) and 
GEMINI invokes the model of sequential binary decay ( SBD ). 
An enhancement of the yields of n-rich nuclei over the prediction of such 
calculations was observed in the reaction $^{86}$Kr+$^{64}$Ni 
at 25 AMeV \cite{GSKrNi}. The shapes of the velocity spectra suggested a 
process with a short timescale such as the very peripheral collisions 
where the details of neutron and proton density profiles 
at the projectile and target surfaces can play a significant role. 
An attempt was made in \cite{GSKrSn} to implement the density-dependent 
correction into the DIT model and an improvement of agreement 
with experimental data from the reactions 
$^{86}$Kr+$^{64}$Ni,$^{112,124}$Sn was observed for some regions, which 
however was not consistent enough to be conclusive and a need for further 
development of a self-consistent density-dependent DIT model 
or an equivalent approach was inferred. 

\section*{ Model analysis of heavy residue data }

\subsection*{Reaction $^{86}$Kr+$^{64}$Ni}

\begin{figure}[h]
\centering
\vspace{5mm}
\includegraphics[width=11.5cm,height=7.75cm]{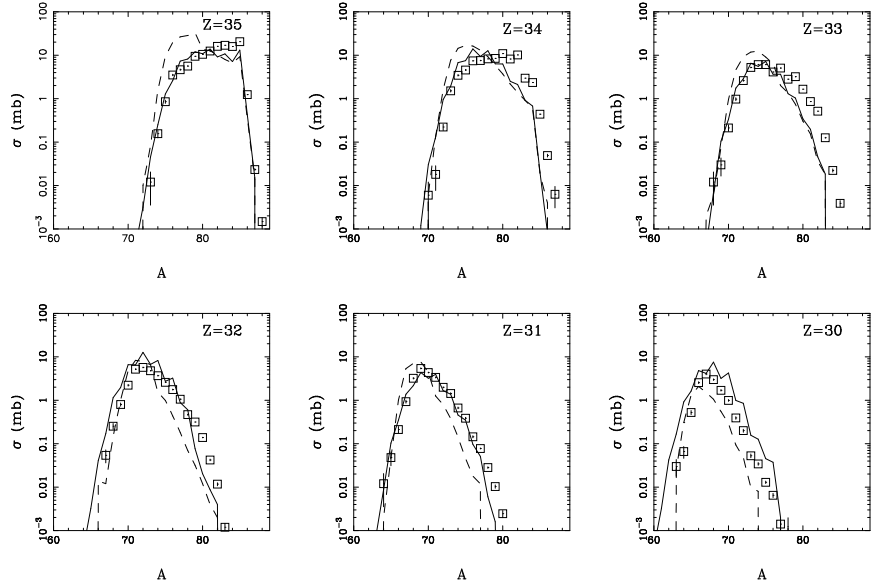}
\caption{\footnotesize
Experimental mass distributions ( symbols ) of elements 
with Z = 30 - 35 observed in the reaction $^{86}$Kr+$^{64}$Ni at 25 AMeV 
\cite{GSKrNi} 
compared to the results of the standard DIT calculations 
combined with the de-excitation codes GEMINI and SMM ( solid and dashed line, 
respectively ). 
}
\label{fgkrni64}
\end{figure}

In Fig. \ref{fgkrni64} we present the experimental mass distributions 
of elements with Z = 30 - 35 observed within the separator acceptance 
in the reaction $^{86}$Kr+$^{64}$Ni at 25 AMeV 
\cite{GSKrNi} compared to the results of the standard  DIT calculations 
\cite{TG} combined with two de-excitation codes SMM \cite{SMM} ( dashed line ) 
and GEMINI  \cite{GEM} ( full line ), 
used for de-excitation of the quasiprojectiles emerging after the DIT 
stage. The simulated yields were corrected for angular acceptance 
of the separator positioned at $0^{\circ}$ ( covering polar angles 
1.0 - 2.7 $^{\circ}$ \cite{GSKrNi} ). 
One observes an excess of experimental yields of neutron-rich 
nuclei with Z = 34 - 32, which is evident both for the SMM and the GEMINI, 
in the former case the effect appears to extend to lower atomic numbers. 
As suggested in \cite{GSKrNi}, such excessive yields of neutron-rich nuclei 
can be caused by the effect of the neutron-rich surface of the target nucleus, 
which in peripheral collisions can lead to stronger flow of neutrons 
from the target to the projectile 
( or flow of protons in the opposite direction ), thus reverting the flow 
toward isospin equilibration. 

\begin{figure}[h]
\centering
\vspace{5mm}
\includegraphics[width=13.cm,height=7.cm]{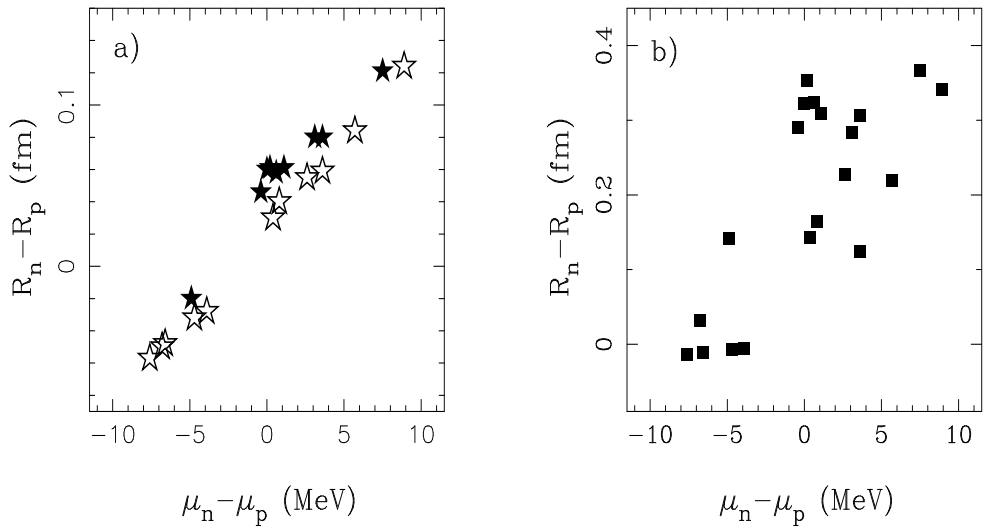}
\caption{\footnotesize
a) Calculated dependence of $R_n-R_p$ on the difference 
of neutron and proton chemical potentials $\mu_n-\mu_p$, obtained using 
the extended Thomas-Fermi model \cite{ETF}. 
Open and solid symbols represent the nuclei with A$<$100 and A$>$100, 
respectively. 
b) As in a), but the values of $R_n-R_p$ are calculated using 
the liquid-droplet parametrization \cite{TG,LDSkin}. 
}
\label{fgcncp}
\end{figure}

In order to understand how the properties of 
nuclear surface can revert the isospin flow one first needs to clarify 
whether and to what extent such effect of nuclear surface is implemented 
in the standard DIT model of Tassan-Got \cite{TG}. The model of Tassan-Got 
is based on various macroscopic parametrizations. For an estimate of 
the thickness of neutron-skin a parametrization based on liquid-droplet model 
of ref. \cite{LDSkin} is used. The values of neutron-skin thickness $t$ 
are then used 
to determine the radius of the sharp uniformly charged sphere which is used 
to describe the single-particle Coulomb potential \cite{TG}. Furthermore, 
the radius of the nuclear part of single-particle potential is additionally 
enhanced by 0.5 fm in order to match the typical radius of the proximity 
potential \cite{Prox}, which allows to describe the reaction dynamics properly. 
Thus such adjustement essentially leads to further enhancement of 
the thickness of neutron-rich surface. However, 
the relation between neutron and proton single-particle potentials 
and corresponding density distributions, on which the nuclear surface is 
usually defined, is not simple and can be, in principle, determined 
only in self-consistent manner. 
In any case, the DIT model of Tassan-Got \cite{TG} proved 
successful in describing the bulk N/Z-equilibration and thus explanation 
to the enhancement observed in Fig. \ref{fgkrni64} will be apparently related 
to the effects not considered in it, specifically these which will lead 
to modification of isospin-asymmetry at nuclear periphery. Also it is 
of interest to identify observables characterizing such effects. 

The observable which is of particular interest is the isospin asymmetry 
at the nuclear periphery, specifically its deviation from the bulk isospin 
asymmetry of the nucleus. It can be shown that the value of this observable 
at the mean half-density radius $(R_n+R_p)/2$ is determined dominantly 
by its linear correlation to the the thickness of a neutron-rich 
surface $R_n-R_p$. An estimate of the thickness of 
a neutron-rich surface $R_n-R_p$, where $R_n$, $R_p$ are the half-density 
radii of neutrons and protons, can be obtained using nuclear structure 
calculations. A calculated dependence of $R_n-R_p$ on the difference 
of neutron and proton chemical potentials $\mu_n-\mu_p$, obtained using 
the extended Thomas-Fermi code of Kolomietz et al. \cite{ETF}, is presented 
in Fig. \ref{fgcncp}, along with the liquid-droplet estimate 
$R_n-R_p = 2.714((1 - 2 Z/A)-0.0016582 Z/A^{1/3})/(1+3.45/A^{1/3})$ 
\cite{TG,LDSkin}. 
The calculations were performed for the $\beta$-stable nuclei 
with masses A = 16 - 238 ( representing typical target nuclei ) with  
the neutron-rich isotopes $^{94}$Kr, $^{132}$Sn being added in order to extend 
the N/Z-range. For the extended Thomas-Fermi calculation, 
one observes a linear correlation for both subsets of nuclei, 
for heavier nuclei ( solid symbols ) the values of $R_n-R_p$ appear 
to be somewhat larger than for lighter nuclei ( open symbols ), 
thus implying a thicker neutron-rich surface for heavier $\beta$-stable nuclei. 
The values obtained using the liquid-droplet parametrization ( squares ) follow 
similar trend, however the absolute values are significantly larger than in 
extended Thomas-Fermi calculation. The spread of the values is caused by 
the use of $\mu_n-\mu_p$ from the previous plot, which do not follow 
the trend of isospin asymmetry so strictly as it is assumed in the 
liquid-drop model, where both $R_n-R_p$ and $\mu_n-\mu_p$ 
follow the isospin asymmetry linearly. The global correlation of 
$R_n-R_p$ with $\mu_n-\mu_p$ suggests the possibility to relate the effect of 
isospin asymmetry at the surface to the difference of chemical potentials. 
However, it should be noted that the values of $\mu_n-\mu_p$ calculated 
using the extended Thomas-Fermi model represent the "smooth" ( macroscopic ) 
nuclear properties and may differ from the experimental values 
of the difference of neutron and proton separation energies $-(S_n-S_p)$ 
which is typically used as the estimate of $\mu_n-\mu_p$ for real nuclei 
at ground state. On the other hand, when assuming that the correlation 
of $R_n-R_p$ with $\mu_n-\mu_p$ is preserved also for real nuclei, the  
experimental values of $-(S_n-S_p)$ can be assumed as representing the 
values of $R_n-R_p$ of such nuclei, with the density profiles modified 
due to the microscopic structure, including the effects not considered 
in the liquid-droplet parametrization used by DIT model 
of Tassan-Got such as shell structure. 

\begin{figure}[h]
\centering
\vspace{5mm}
\includegraphics[width=11.5cm,height=7.75cm]{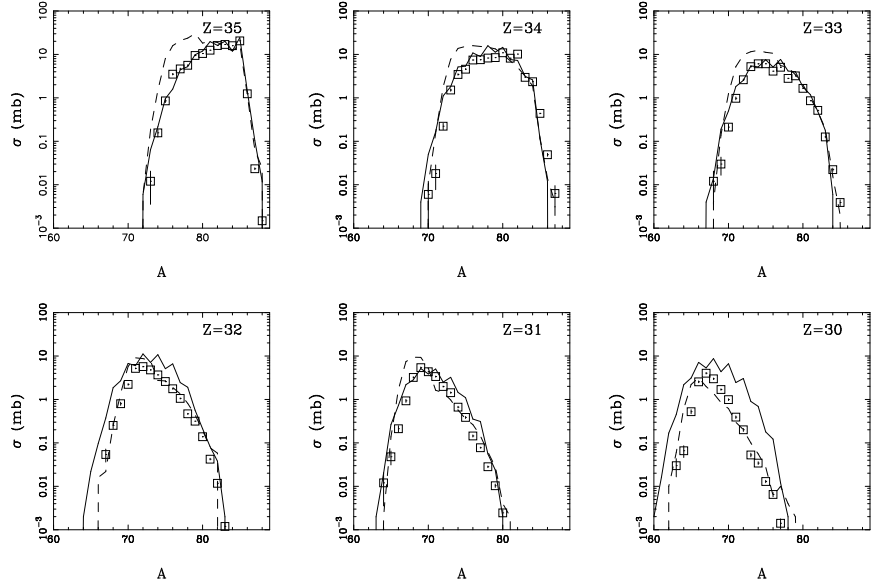}
\caption{\footnotesize
Experimental mass distributions ( symbols ) of elements 
with Z = 30 - 35 observed in the reaction $^{86}$Kr+$^{64}$Ni at 25 AMeV 
\cite{GSKrNi} 
compared to the results of the modified DIT calculations 
combined with the de-excitation codes GEMINI and SMM ( solid and dashed line, 
respectively ). 
}
\label{fgkrni64m}
\end{figure}

The correlation of the $R_n-R_p$ with $\mu_n-\mu_p$ shown in Fig. \ref{fgcncp} 
suggests the possibility to implement a corresponding correction into the 
DIT model taking into account the effect of nuclear periphery.  
Based on the discussion of Fig. \ref{fgcncp} in the previous paragraph, 
in particular considering the possibility to estimate the surface 
properties of real nuclei using the value of $-(S_n-S_p)$, 
we made a minor modification in the 
DIT code of Tassan-Got \cite{TG} by scaling the transfer probabilities 
by the exponential factors

\begin{flushleft}
\begin{eqnarray}
P_n (P \rightarrow T) & \longrightarrow &
{\rm e}^{-0.5\kappa(\delta S_{nP}-\delta S_{pP}
-\delta S_{nT}+\delta S_{pT})} P_n (P \rightarrow T) \nonumber \\
P_p (P \rightarrow T) & \longrightarrow &
{\rm e}^{0.5\kappa(\delta S_{nP}-\delta S_{pP}
-\delta S_{nT}+\delta S_{pT})} P_p (P \rightarrow T) \nonumber \\
P_n (T \rightarrow P) & \longrightarrow &
{\rm e}^{-0.5\kappa(\delta S_{nT}-\delta S_{pT}
-\delta S_{nP}+\delta S_{pP})} P_n (T \rightarrow P) \nonumber \\
P_p (T \rightarrow P) & \longrightarrow &
{\rm e}^{0.5\kappa(\delta S_{nT}-\delta S_{pT}
-\delta S_{nP}+\delta S_{pP})} P_p (T \rightarrow P)
\label{snsp}
\end{eqnarray}
\end{flushleft}

where $\kappa$ is a free parameter ( due to essentially a Boltzman factor 
structure, possibly representing an inverse statistical temperature ) and 
$\delta S_{nP}$, $\delta S_{pP}$, $\delta S_{nT}$, 
$\delta S_{pT}$ represent 
the differences of neutron and proton separation energies for the projectile 
and target calculated using the experimental \cite{AuW} and liquid-drop 
\cite{MySw} masses, thus expressing the effect of the microscopic structure. 
The smooth part is subtracted from the experimental values due to the fact 
that the macroscopic values of $\mu_n-\mu_p$ follow the bulk N/Z-ratios 
of reaction partners and the bulk N/Z equilibration is described consistently 
by the DIT code of Tassan-Got. Thus, the DIT model is supplemented with 
phenomenological information 
on shell structure at the nuclear periphery which can explain 
the deviation of nucleon exchange from the path toward isospin 
equilibration, and the model framework assumes the structure commonly used e.g. 
to describe the ground state properties of nuclei and fission \cite{Strut}, 
where the liquid-drop model ( or extended Thomas-Fermi model in more 
recent works as for instance in ref. \cite{ETFSI} ) 
is supplemented by additional term describing the effect 
of shell structure. In the present case, such a term is phenomenologically 
related to shell corrections in neutron and proton separation energies 
$\delta S_{nP}$, $\delta S_{pP}$, $\delta S_{nT}$, $\delta S_{pT}$, 
which are based on experimental information and thus their use can 
result in enhanced predictive power.
The modified DIT calculation was used only for 
non-overlapping projectile-target configurations, consistent with 
the assumption that it represents an effect of nuclear periphery.   
A cut-off was set at zero half-density surface separation 
( representing touching half-density surfaces ) below which 
a standard DIT calculation, following the path toward isospin equilibration, 
was used. 

\begin{figure}[h]
\centering
\vspace{5mm}
\includegraphics[width=11.5cm,height=7.75cm]{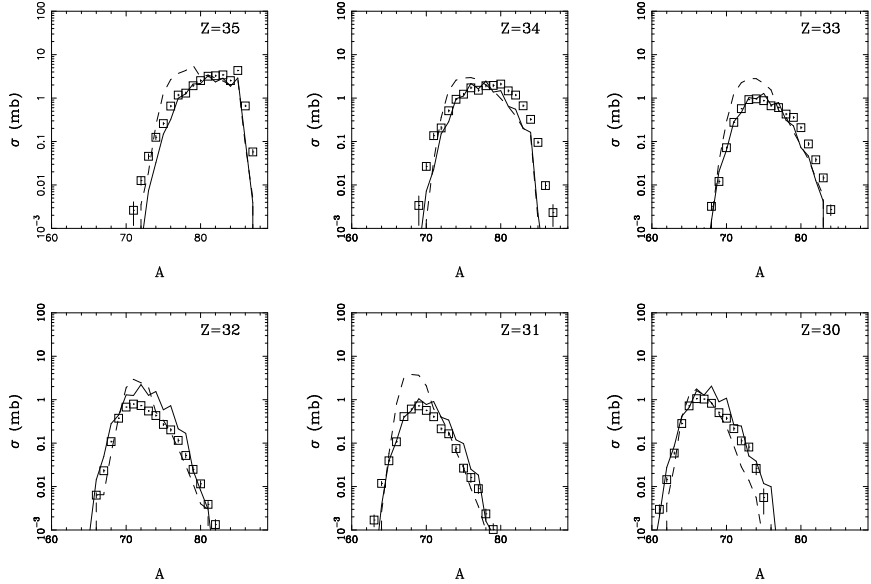}
\caption{\footnotesize
Experimental mass distributions ( symbols ) of elements 
with Z = 30 - 35 observed in the reaction $^{86}$Kr+$^{124}$Sn at 25 AMeV 
\cite{GSKrSn} 
compared to the results of the standard DIT calculations 
with GEMINI ( full line ) and SMM ( dashed line ).   
}
\label{fgkrsn24}
\end{figure}
 
In Fig. \ref{fgkrni64m} we present the experimental mass distributions 
of elements with Z = 30 - 35 observed within the separator acceptance 
in the reaction $^{86}$Kr+$^{64}$Ni at 25 AMeV 
\cite{GSKrNi} compared to the results of the modified DIT calculations,  
again combined with the two de-excitation codes GEMINI  \cite{GEM}
( full line ) and SMM \cite{SMM} ( dashed line ). 
The simulated yields were filtered for angular acceptance 
as in the case of Fig. \ref{fgkrni64}.   
Several calculations were performed with different values of $\kappa$ and  
the value of $\kappa$ = 0.53, used in the modified DIT calculations 
presented in Fig. \ref{fgkrni64m}, was obtained as an optimum value 
best reproducing the experimental mass distributions. 
One can observe that the modification of the DIT code allows 
to dramatically enhance the agreement with the experimental yields of 
neutron-rich nuclei with Z = 35 - 32 when using both de-excitation codes.  
The GEMINI calculation results in the nearly symmetric mass distributions 
which appear to overestimate the widths of mass distributions 
of lighter elements. The SMM calculation appears to reproduce well 
the yields of neutron-rich nuclei also for lighter elements, on the other 
hand the yields of $\beta$-stable isotopes appear to be overestimated.  
Such differences of GEMINI and SMM calculations are in good agreement 
with the results of the work \cite{MVSnAl} where the SMM calculations lead 
to better reproduction of yields originating from the hot quasiprojectiles, 
while for the colder quasiprojectiles with excitation energies 1-2 AMeV 
GEMINI performed better due to the implementation of sequential binary 
decay which is missing in the simulation of secondary emission 
from the hot fragments in SMM. 

\begin{figure}[h]
\centering
\vspace{5mm}
\includegraphics[width=11.5cm,height=7.75cm]{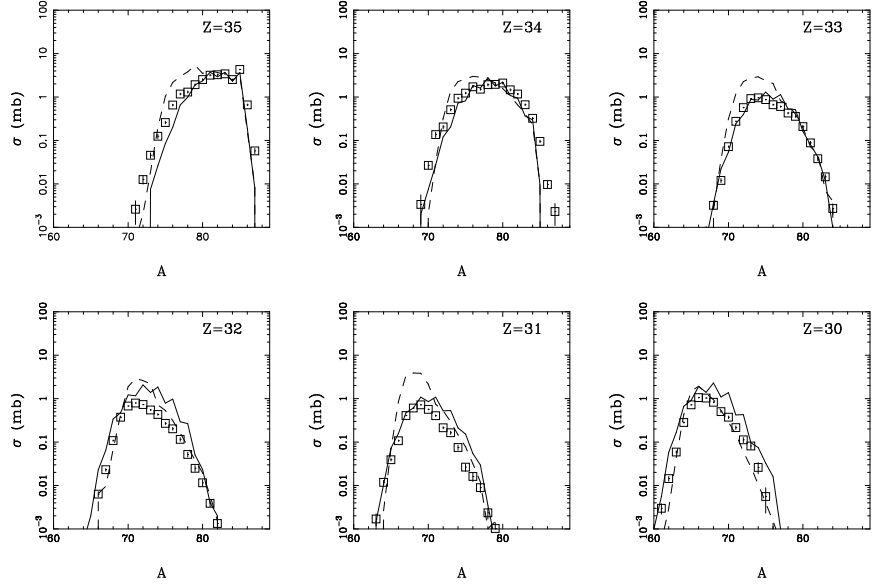}
\caption{\footnotesize
Experimental mass distributions ( symbols ) of elements 
with Z = 30 - 35 observed in the reaction $^{86}$Kr+$^{124}$Sn at 25 AMeV 
\cite{GSKrSn} 
compared to the results of the modified DIT calculations 
with GEMINI ( full line ) and SMM ( dashed line ).   
}
\label{fgkrsn24m}
\end{figure}
 
\subsection*{Reactions $^{86}$Kr+$^{112,124}$Sn}

In order to further examine the modification 
of the DIT code, the results of simulations were compared 
to data from other reactions. Figs. \ref{fgkrsn24}, \ref{fgkrsn24m} 
show the experimental mass distributions of elements 
with Z = 30 - 35 observed within the separator acceptance 
in the reaction $^{86}$Kr+$^{124}$Sn at 25 AMeV 
\cite{GSKrSn} again compared to the results of the standard 
( Fig. \ref{fgkrsn24} ) and modified ( Fig. \ref{fgkrsn24m} ) 
DIT calculations combined with the GEMINI ( full line ) and
SMM ( dashed line ). 
The simulated yields were filtered for angular acceptance 
of the separator positioned at $4^{\circ}$ ( covering polar angles 
2.7 - 5.4 $^{\circ}$ \cite{GSKrSn} ) with appropriate azimuthal 
corrections. One can see that in this case, 
the overall agreement with the experimental data is improved 
in the modified DIT calculation ( using $\kappa=0.53$ as in the previous case ) 
when comparing to the standard one. 
As in the previous case, the modified DIT calculation with SMM appears 
to reproduce the shapes of mass distributions more consistently 
( except the overestimation of the yields of $\beta$-stable nuclei ), 
while the GEMINI code appears to lead to more symmetric mass 
distributions with overestimated width at lower atomic numbers. 

\begin{figure}[h]
\centering
\vspace{5mm}
\includegraphics[width=11.5cm,height=7.75cm]{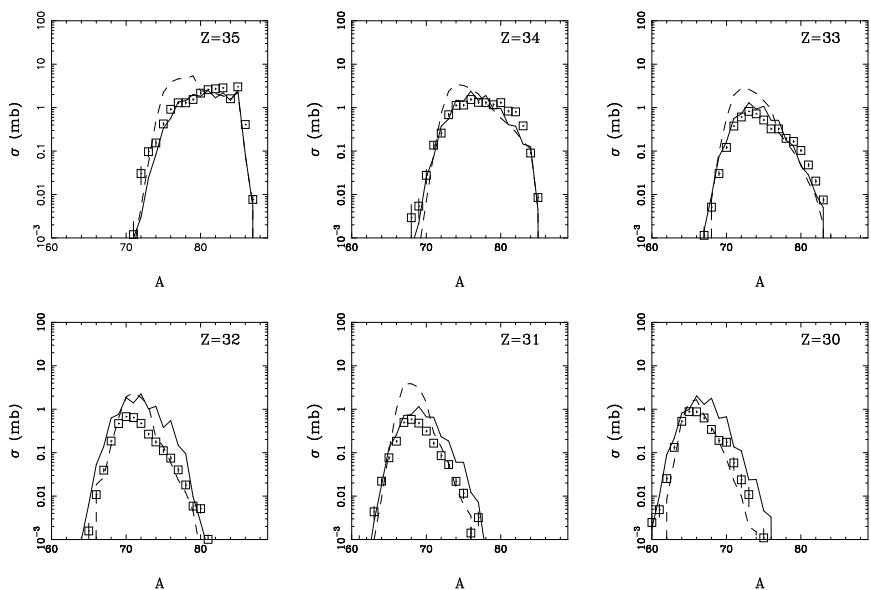}
\caption{\footnotesize
Experimental mass distributions ( symbols ) of elements 
with Z = 30 - 35 observed in the reaction $^{86}$Kr+$^{112}$Sn at 25 AMeV 
\cite{GSKrSn} 
compared to the results of the standard DIT calculations 
with GEMINI ( full line ) and SMM ( dashed line ).   
}
\label{fgkrsn12}
\end{figure}
 
\begin{figure}[h]
\centering
\vspace{5mm}
\includegraphics[width=11.5cm,height=7.75cm]{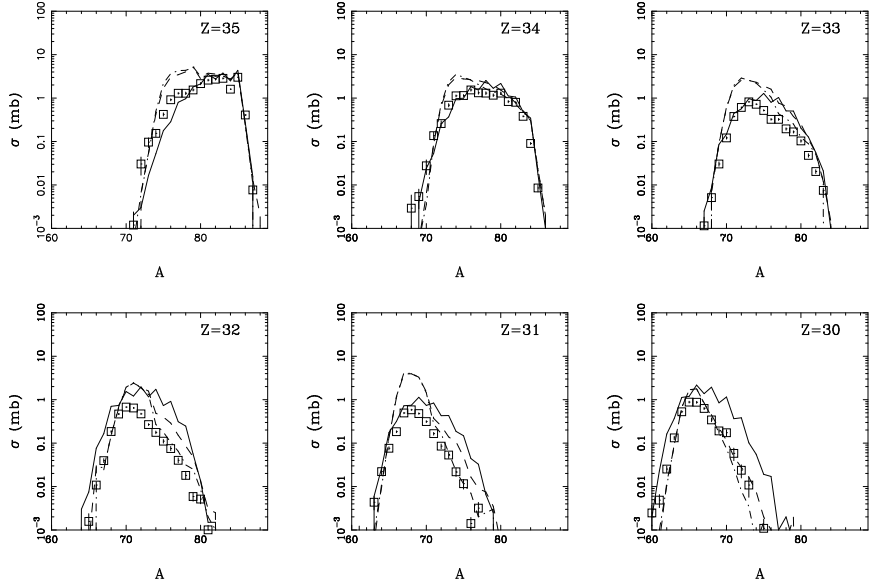}
\caption{\footnotesize
Experimental mass distributions ( symbols ) of elements 
with Z = 30 - 35 observed in the reaction $^{86}$Kr+$^{112}$Sn at 25 AMeV 
\cite{GSKrSn} 
compared to the results of the modified DIT calculations 
with GEMINI ( full line ) and SMM ( dashed line ). Dash-dotted line 
represents the modified DIT with a cutoff shifted to 0.8 fm combined with SMM.    
}
\label{fgkrsn12m}
\end{figure}

Figs. \ref{fgkrsn12}, \ref{fgkrsn12m} present the experimental mass 
distributions of elements with Z = 30 - 35 observed within the separator 
acceptance in the reaction $^{86}$Kr+$^{112}$Sn at 25 AMeV 
\cite{GSKrSn}, again compared to the results of the simulations,  
the standard ( Fig. \ref{fgkrsn12} ) and modified ( Fig. \ref{fgkrsn12m} ) 
DIT calculations combined with the GEMINI ( full line ) and
SMM ( dashed line ). 
The simulated yields were filtered for angular acceptance 
of the separator positioned at $4^{\circ}$ ( as in the case of 
reaction $^{86}$Kr+$^{124}$Sn ). 
In this case the modified DIT calculation ( using the value $\kappa=0.53$ 
successful in previous cases ) combined with 
GEMINI leads to improvement for Z = 35 - 34, consistent with previous 
cases. However, both the modified and, to a lesser extent, the standard 
DIT calculation combined with GEMINI appear to overestimate the 
yields of neutron-rich nuclei with Z = 33 - 30 due to shifted centroids 
and overestimation of both maximum value and width. The modified 
DIT calculation combined with SMM appears to reproduce the shapes 
of mass distribution consistent to previous cases, well at the 
neutron-rich side ( specifically for the Z = 35 - 34 ), while overestimating 
the yields of $\beta$-stable 
nuclei. For Z = 33 - 31 the overall elemental yields appear to be overestimated 
( even when subtracting the excess yields of $\beta$-stable nuclei ) 
and the standard DIT calculation combined with SMM 
reproduces these yields better. Such a situation in the case 
of the reaction $^{86}$Kr+$^{112}$Sn 
may signal that an increase of proton transfer probability into such 
a massive proton-rich target according to Eqn. (\ref{snsp}) 
may be reverted at smaller separation distances 
by the effect of increasingly repulsive Coulomb interaction.  
The dash-dotted line in Fig. \ref{fgkrsn12m} represents a modified 
DIT calculation ( $\kappa=0.53$ ) where the cut-off is shifted to minimal 
separation of half-density surfaces equal to 0.8 fm, again combined with SMM. 
Such a calculation 
reproduces the experimental mass distributions much better, thus suggesting 
that for the proton-rich target $^{112}$Sn, the proton transfer barrier 
assumes its sensitivity to isospin asymmetry of nuclear periphery 0.8 fm 
outwards when compared to neutron-rich targets, due to the effect of stronger 
Coulomb repulsion at more compact di-nuclear configurations than 
predicted by the approximation used. 

\section*{Conclusions}

In summary, the DIT model of Tassan-Got \cite{TG} is supplemented 
with a phenomenological correction introducing  
the effect of shell structure on nuclear periphery.  
A consistent agreement with experimental data is achieved in the reactions 
of a 25 AMeV $^{86}$Kr beam with three different target nuclei, 
specifically allowing to describe the deviation of the nucleon 
exchange from the path toward isospin equilibration. 
The value of parameter $\kappa$ appears to be system-independent, 
consistent with the system-independent correlation of neutron skin 
thickness with difference of neutron and proton chemical potentials. 
The success of the modified DIT calculation can be 
explained as a correction reflecting the modification of 
neutron and proton transfer probabilities, most probably of the transfer 
barriers, in peripheral collisions due to the effect of shell structure 
on isospin asymmetry at nuclear periphery. 
Such an effect is of interest for better prediction of the production 
rates of exotic nuclei with a wide range of N/Z-ratios 
at the new generation of rare isotope beam facilities. 
Discrepancies observed for the most proton-rich target nucleus $^{112}$Sn 
signal a loss of sensitivity of the transfer barriers toward 
isospin asymmetry at the nuclear periphery due to increased 
Coulomb repulsion at non-overlapping di-nuclear configurations 
with surface separation below 0.8 fm. 
The de-excitation code GEMINI appears to systematically lead 
to overly symmetric mass distributions with increasingly 
overestimated widths toward the lighter elements. 
The SMM appears to reproduce the experimental shapes better 
( except for excess yields of $\beta$-stable isotopes ).  
In particular the trends of the yields of neutron-rich nuclei are reproduced 
consistently. 

This work was supported through grant of Slovak Scientific Grant Agency
VEGA-2/5098/25 and by the Department of Energy through 
grant No. DE-FG03-93ER40773.

\end{document}